\documentclass[conference,a4paper]{Latex-2025/APSIPA2025}
\usepackage{amsmath}
\usepackage{graphicx}
\usepackage{multirow}
\usepackage{threeparttable}
\usepackage[backend=biber,style=ieee,]{biblatex}
\usepackage{amssymb}
\usepackage{booktabs}

\usepackage{geometry}
\usepackage{fancyhdr}

\begin{document}

\title{ASCMamba: Multimodal Time-Frequency Mamba for Acoustic Scene Classification}

\author{
\authorblockN{
Bochao Sun\authorrefmark{1} and
Dong Wang\authorrefmark{2} and
Zhanlong Yang\authorrefmark{1} and
Jun Yang\authorrefmark{2} and
Han Yin\authorrefmark{3}
}

\authorblockA{
\authorrefmark{1}
School of Marine Science and Technology, Northwestern Polytechnical University, Xi’an, China \\
\authorrefmark{2}
School of Automation, Northwestern Polytechnical University, Xi’an, China    \\
\authorrefmark{3}
School of Electrical Engineering, KAIST, Daejeon, Republic of Korea \\
}
}

\maketitle

\begin{abstract}
 Acoustic Scene Classification (ASC) is a fundamental problem in computational audition, which seeks to classify environments based on the distinctive acoustic features.
 In the ASC task of the APSIPA ASC 2025 Grand Challenge, the organizers introduce a multimodal ASC task. Unlike traditional ASC systems that rely solely on audio inputs, this challenge provides additional textual information as inputs, including the location where the audio is recorded and the time of recording.
 In this paper, we present our proposed system for the ASC task in the APSIPA ASC 2025 Grand Challenge. 
 Specifically, we propose a multimodal network, \textbf{ASCMamba}, which integrates audio and textual information for fine-grained acoustic scene understanding and effective multimodal ASC. The proposed ASCMamba employs a DenseEncoder to extract hierarchical spectral features from spectrograms, followed by a dual-path Mamba blocks that capture long-range temporal and frequency dependencies using Mamba-based state space models.
 In addition, we present a two-step pseudo-labeling mechanism to generate more reliable pseudo-labels.
 Results show that the proposed system outperforms all the participating teams and achieves a 6.2\% improvement over the baseline. Code, model and pre-trained checkpoints are available at https://github.com/S-Orion/ASCMamba.git

 %achieve superior performance, with an improvement %ranging from 4\% to 5\% over the challenge baseline.
\end{abstract}

\section{INTRODUCTION}
Acoustic scene classification (ASC) is a crucial research problem in computational audition that aims to recognize the unique acoustic characteristics of an environment~\cite{barchiesi2015acoustic,yin2025exploring,yin2025multi}. Potential applications of ASC techniques include environmental monitoring and smart devices. Yet prevailing methods often assume static scenes, neglecting spatiotemporal variability across locations and record times. Ignoring such context undermines model generalization in real-world deployments.

Unlike the ASC task in the ICME 2024 Challenge~\cite{bai2024description}, the APSIPA ASC 2025 Grand Challenge focuses on two critical factors influencing the performance of the ASC task: additional contextual information and scarcity of labeled data. The problem of leveraging additional contextual information, such as city-level location data and precise timestamps, is explored in this challenge. Another key issue is utilizing abundant unlabelled data to train robust ASC systems.

%Recently, a structured state-space model termed Mamba~\cite{gu2023mamba} has emerged. By introducing a selective state mechanism, it reduces the computational complexity of sequence modeling from the quadratic scale of Transformer to a linear one, thereby enabling efficient and powerful modeling of extremely long sequences. 

%we propose \textbf{ASCMamba}, a multi-modal network for ASC tasks.% 

In this paper, we present our approach for the ASC task in the APSIPA ASC 2025 Grand Challenge. Specifically, we adopt Mamba~\cite{gu2023mamba} as the backbone of our model, as it exhibits more prominent efficiency and performance advantages over Transformer~\cite{vaswani2017attention} in processing long-duration audio sequences. To fully exploit the temporal and spectral dependencies in audio signals, ASCMamba applies multiple Mamba blocks for dynamic modeling in both the record time and frequency domains.  Furthermore, to facilitate multi-modal information interaction, we adopt a Conditional Layer Normalization (CLN) mechanism to incorporate text embeddings into ASCMamba.   

The Challenge offers an extensive collection of unlabeled data, which can be leveraged for semi-supervised learning approaches. In this work, we first pre-train the proposed ASCMamba on TAU Urban Acoustic Scenes (UAS) 2020 Mobile development dataset~\cite{toni2020tau} and CochlScene dataset~\cite{jeong2022cochlscene}. The labeled data from Chinese Acoustic Scene (CAS) development dataset to fine-tune the pre-trained ASCMamba. For unlabeled CAS samples, we use the first fine-tuned ASCMamba to generate pseudo labels. For certain unlabeled samples, the pseudo-labels generated by the ASCMamba model have low confidence. To improve the quality of these pseudo-labels, we develop a secondary system dedicated to generating pseudo-labels for above mentioned low-confidence cases, and then use the intersection of these pseudo-labels with those predicted by the ASCMamba model to produce the reliable pseudo-labels. The second system is based on the challenge baseline, i.e., SE-Trans~\cite{9951400}. We improve the SE-Trans architecture by incorporating multi-scale pooling to enhance the ability of feature representation. In addition, we introduce an extra fully connected layer for indoor/outdoor binary classification as a prior. We then adjust the   ASC class confidence scores based on the binary classification results to further improve accuracy. Finally, the ASCMamba is fine-tuned once more on the union of labeled and pseudo-labeled data, which serves as the final ASC model for evaluation. 

\begin{figure*}[t!]
\centering
\includegraphics[width=2\columnwidth]{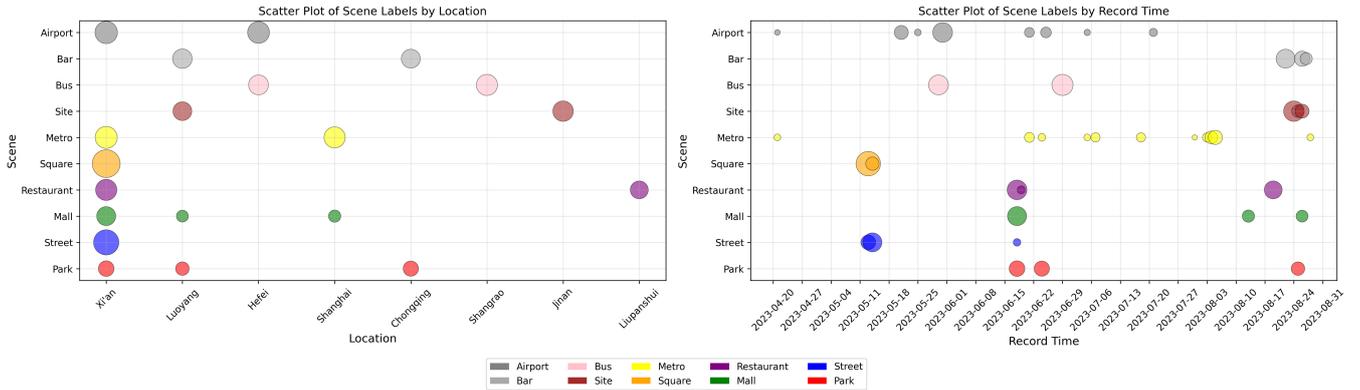}
\caption{Distribution of acoustic scenes in development set by location and record time.}
\label{fig:Distribution}
\end{figure*}
\section{DATASETS}

\subsection{Overview}
The TAU UAS 2020 Mobile development dataset~\cite{toni2020tau} and the CochlScene dataset~\cite{jeong2022cochlscene} serve as the sources for pre-training ASCMamba model. TAU UAS 2020 Mobile comprises 23,046 samples, each delivered in binaural format at a 48 kHz sampling rate. CochlScene provides 76,115 single-channel audio files sampled at 44.1 kHz. Because these datasets cover different acoustic-scene taxonomies, we remove selected scene labels and merge others to construct a unified pre-training set. Table~\ref{tab:dataset} lists the resulting counts of audio recordings per scene, where the data are used to pre-train the proposed ASCMamba model. The CAS 2023 development dataset comprises 8,700 audio clips, 20\% of which are labeled. We extract the labeled data as the initial labeled dataset to fine-tune the ASCMamba and improved SE-Trans models.
% These labeled data are merged with the new pre-training dataset to create the initial labeled dataset.

\begin{table}
    \centering
\caption{The number of audio recordings for each scene in the new generated
}
\label{tab:placeholder}
    \begin{tabular}{cc}
         \multicolumn{2}{c}{Pre-training dataset}\\ \hline
         Scene& Number of audio recordings\\ \hline
         Airport& 2,302\\
         Bus& 8,125\\
         Car& 5,845\\
         Metro& 8,201\\
         Metro Station& 8,201\\
         Public square& 2,303\\
         Restaurant& 5,933\\
         Shopping Mall& 2,303\\
 Traffic Street&8,049\\
 Urban Park&8,048\\ \hline
 Total&59,310\\ \hline
    \end{tabular}
    \label{tab:dataset}
\end{table}

%The CAS 2023 development dataset comprises 8,700 audio clips, 20\% of which are annotated. These labeled data are merged with the new pre-training dataset to create the initial labeled dataset.
% These labeled data and the new pre-training dataset will be used as the initial labeled dataset.
% These labeled data will be used as the initial labeled dataset to fine-tune the ASCMamba model.
% The labeled data serve as the initial dataset for fine tuning both our proposed ASCMamba model and improved SE-Trans moedl.

\subsection{Location and Record Time Distribution Analysis}
In the development dataset, the number of labeled samples is 1,740, and the spatial and temporal distribution of scene labels exhibit significant characteristics. From a spatial perspective, the development dataset comprises data from 8 locations, including \textit{Xi'an}, \textit{Luoyang}, \textit{Hefei}, \textit{Shanghai}, \textit{Chongqing}, \textit{Shangrao}, \textit{Jinan}, and \textit{Liupanshui} in descending order of representation. The distribution of scene labels over location and record time is shown in Fig.~\ref{fig:Distribution}. The size of the circle represents the density of scene labels, and a larger circle indicates a higher number of scene labels for a specific location or record time period. As can be seen, the spatial and temporal information exhibit a strong correlation with scene labels. This provides an important basis for utilizing spatiotemporal information as context for the classification task.  

It should be noted that there are 12 locations in the evaluation set, with 6 overlapping in the development dataset and the remaining 6 being exclusive to the evaluation dataset. Regarding temporal coverage, the development dataset spans from April 21, 2023, to August 28, 2023, while the evaluation dataset spans from April 22, 2023, to August 27, 2023.

% \begin{figure}
% \centering
% \includegraphics[width=\linewidth]{Latex-2025/组合散点图.png}
% \caption{Scatter plot of the distribution of scene labels by location and time.}
% \label{fig:Scatter}
% \end{figure}

\begin{figure*}[t!]
\centering
\includegraphics[width=1.6\columnwidth]{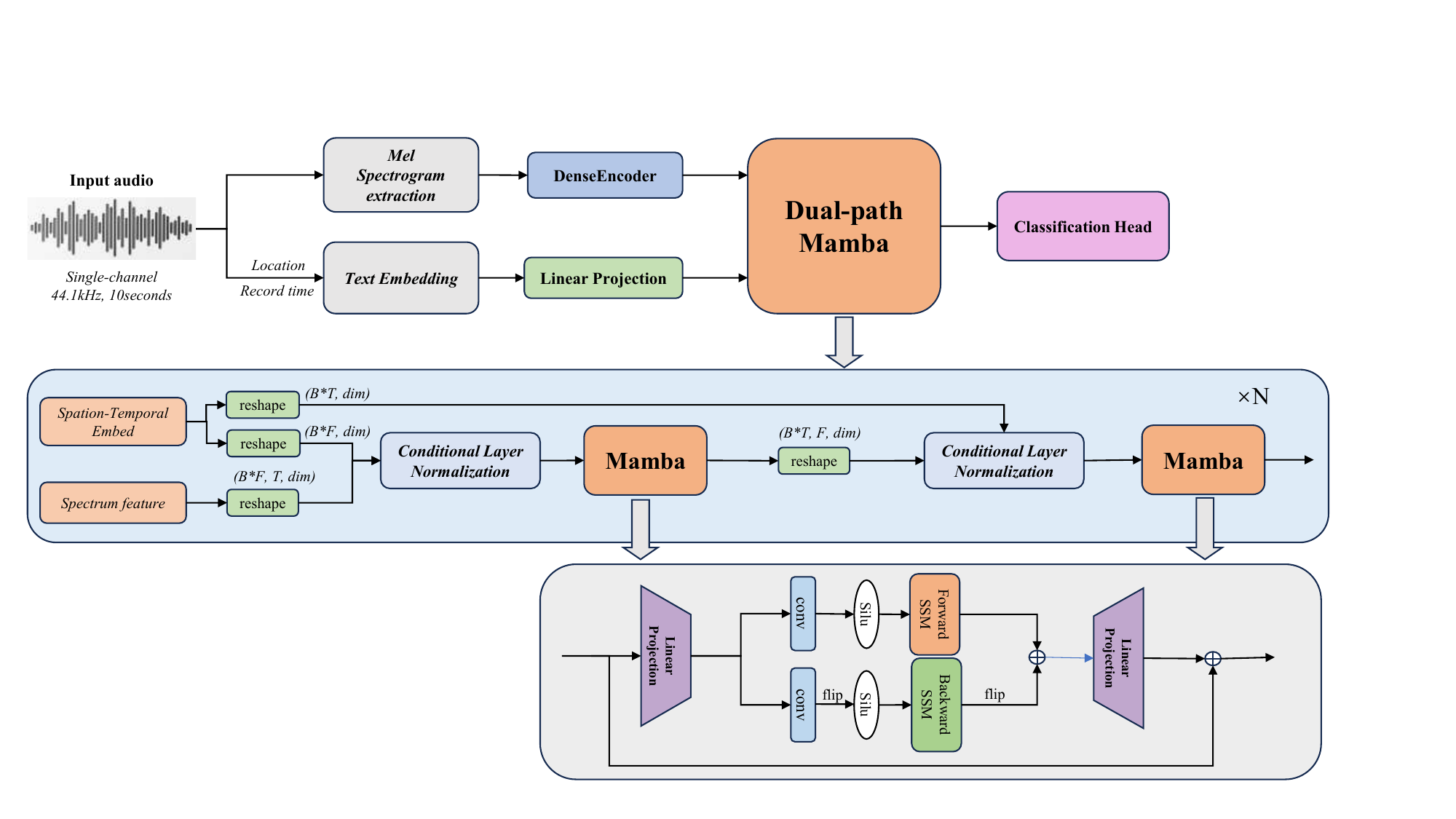}
\caption{Overview of the proposed ASCMamba, which is composed of a DenseEncoder and a Dual-path Mamba Block.}
\label{fig:overview}
\end{figure*}

\section{PROPOSED APPROACH}

\subsection{ASCMamba}

As shown in Fig~\ref{fig:overview}, the proposed ASCMamba model is composed of 2 blocks: DenseEncoder and Dual-path Mamba Block.
Details are described as follows.

\subsubsection{DenseEncoder}
The DenseEncoder is a two-dimensional convolutional feature extraction module designed for time-frequency representation learning in audio processing tasks. It consists of three primary components: an initial channel projection block, a dense connectivity-based feature refinement block, and a frequency-axis down sampling block. Specially, a DenseBlock with a depth of 4 is applied to encourage feature reuse and gradient propagation across layers. Inspired by DenseNet~\cite{kim2024densenets}, each layer within the block receives feature maps from all preceding layers as input, promoting the learning of compact and discriminative spectral patterns. It mainly serves as an efficient front-end feature extractor to effectively capture local and hierarchical features in the spectrogram.

\subsubsection{Dual-path Mamba}
The core of the block is a dual-path Mamba architecture, which separately models dynamics along the time and frequency dimensions. Given an input spectrogram feature, which is the encoded representation extracted by the preceding DenseEncoder from log-mel inputs. We denote this feature as $\boldsymbol{X} \in {R}^{B \times C \times T \times F }$, which is reshaped into 2 sequences, i.e., $\boldsymbol{X}_{t}$ and $\boldsymbol{X}_{f}$, as shown in Table~\ref{tab:sequence}. Each sequence is processed independently by a MambaBlock, capturing long-range dependencies along their respective axes. 

\begin{table}
    \centering
\caption{The reshaped sequences}
\label{tab:placeholder}
    \begin{tabular}{cc} \hline
         Sequences& Means\\ \hline 
         $\boldsymbol{X}_{t} \in {R}^{(B * F) \times T \times C}$& each frequency bin as a sequence over record time\\
         $\boldsymbol{X}_{f} \in {R}^{(B * T) \times F \times C}$& each frame as a sequence over frequency\\ \hline 
    \end{tabular}
    \label{tab:sequence}
\end{table}

To enable \textbf{multimodal integration}, the model accepts text embeddings based on the input location and record time information, which are projected into a shared conditional space ${R}^{D_{cond} }$. This conditional vector is used to modulate the internal feature representations through a CLN mechanism. Specifically, CLN dynamically adjusts the affine parameters (scale $\gamma $ and $\beta $ bias ) of layer normalization based on the context, which can be formulated as:

\begin{equation}
    CLN(x, c) = \gamma (c) \cdot LN(x)+\beta (c)
    \label{Eq:1}
\end{equation}

where $\gamma (x)$ and $\beta (x)$ are generated by linear projections from the conditional vector c. This modulation is applied before both the temporal and frequency Mamba paths, allowing the model to adapt its feature space according to spatiotemporal contexts, such as emphasizing different frequency patterns depending on the time of day or geographic region.

\begin{table*}[htbp]
\centering
\caption{The ACC (\%) of baseline and proposed systems on Valid-Easy and Valid-Hard. ``L\&RT'' means ``Location and Record Time''. }
\renewcommand\arraystretch{1}{
\setlength{\tabcolsep}{0.9mm}{
\begin{tabular}{c|cccccccccc|c}
\toprule
System & Airport & Bar & Bus & Construction Site & Metro & Republic Square & Restaurant & Shopping Mall & Traffic Street & Urban Park & Average \\

\midrule
\multicolumn{12}{l}{\textit{Valid-Easy}} \\
\midrule
SE-Trans (Baseline) & 76.92  & 94.59 & 96.88 & 97.06  & 90.91 & 97.44 & 93.33 & 88.89 & 100 & 91.30 & 92.73\\ 
ASCMamba w/ L\&RT &97.07 &100 & 100 & 100 & 90.01 & 100 & 93.01 & 100 & 97.00 & 100 & \textbf{97.71}\\
ASCMamba w/o L\&RT &84.62&91.30 &95.15&94.12 &96.08&93.33&93.33 &92.73 &96.88&100 & 93.75\\
\midrule
\multicolumn{12}{l}{\textit{Valid-Hard (5\%)}} \\
\midrule
ASCMamba w/ L\&RT &82.61& 93.17& 96.15 & 94.87& 97.10 & 95.25& 93.26 & 93.00& 96.00& 100 & 94.14\\
ASCMamba w/o L\&RT &84.62 & 100 & 97.73 & 90.00 & 97.44 & 100 & 94.33 & 100 & 100 & 100 & \textbf{96.41}\\
\bottomrule
\end{tabular}
}
}

\label{tab:result}
\end{table*}

\subsection{Improved SE-Trans}
In order to make full use of the official development dataset, we develop a second system dedicated to generating pseudo-labels for the data with low confidence predicted by the ASCMamba model, and then use the intersection of these pseudo-labels with the labels predicted by the ASCMamba model as the reliable pseudo-labels. The second system adopts an \textbf{improved SE-Trans} architecture to enhance the ability of feature expression and the accuracy of classification through multi-scale pooling and two-step classification strategy.

Specifically, the improved SE-Trans uses 2 Squeeze-and-Excitation (SE) modules (with 2 convolutional layers, a multi-scale SE layer, and pooling), where the SE layers apply 1×1, 2×2, and 3×3 multi-scale pooling to strengthen feature representation. Features are then processed by a Transformer encoder to model spatiotemporal dependencies, followed by 2 fully connected layers outputting 10 specific scene labels and 2 rough labels. The two-class fully connected layer aims to classify an input audio clip into 1 of 2 main classes, including in-door and out-door. The final prediction of scene class is obtained by score fusion of these 2 classifiers~\cite{hu2021two}, which is expressed as follows:

\begin{equation}
    c=argmax_{c,i\supset c} y_{c}^{1}\ast y_{i}^{2}  
    \label{Eq:2}
\end{equation}

where $y_{c}^{1}$ denotes the probability of class \textit{c} predicted by the ten-class classifier, while $y_{i}^{2}$ represents the probability of class \textit{i} predicted by the binary classifier, with $c\in \left \{ 1, 2,...10 \right \} $ , and $i \left \{ 1, 2 \right \} $. Since $i\supset c$, means that class \textit{i} is a superset of class \textit{c}. For instance, the indoor scene category is a superset that includes bus, metro, restaurant, shopping mall and bar. This design significantly improves recognition accuracy and robustness in complex scenarios.

% \begin{table}
%     \centering
% \caption{The performances of the baseline system and the proposed system for each scene class on the validation set in our generated labeled dataset}
% \label{tab:placeholder}
%     \begin{tabular}{ccc} \hline
%          Scene&  Baseline Accuracy& Our model Accuracy\\ \hline
%          Airport&  80.81\%& 100.00\%\\
%          Bar&  97.62\%& 100.00\%\\
%          Bus&  96.07\%& 100.00\%\\
%          Construction Site&  95.24\%& 100.00\%\\
%          Metro&  92.47\%& 98.39\%\\
%          Republic Square&  94.44\%& 100.00\%\\
%          Restaurant&  95.15\%& 100.00\%\\
%          Shopping Mall&  82.42\%& 98.79\%\\
%          Traffic Street&  97.90\%& 99.30\%\\
%  Urban Park& 93.13\%&100.00\%\\ 
%  Average& 92.31\%&99.65\%\\ \hline
%     \end{tabular}
% \end{table}

\subsection{Two-step Pseudo-labeling}

To exploit the unlabeled data, we introduce a two-step pseudo-labeling scheme. In the first step, the pre-trained ASCMamba model is fine-tuned on the initial labeled dataset and subsequently used to assign pseudo labels to the unlabeled clips. The predicted posterior probabilities of unlabeled data are sorted from high to low, and we select the top 90\% of the pseudo-labeled data. In the second step, the ASCMamba and the improved SE-Trans are employed to generate pseudo labels for the left 10\% of the unlabeled data in the development set. Audio samples predicted to belong to the same scene category by both of these 2 models are selected as reliable pseudo-labeled data. These samples are finally combined with the initial labeled dataset to form the definitive labeled dataset used to train our submission system.

\section{Experimental Setups}

\subsection{Validation Sets Creation}
\label{sec:validset creation}

To explore the performance of the baseline and proposed system, we first analyze the distribution discrepancy between the validation set and the training set, and find that their distributions are generally consistent. However, since this competition requires participants to pay attention to the impact of domain shift on audio classification, we divide the validation set into \textbf{Valid-Easy} and \textbf{Valid-Hard} with the help of the official evaluation set to simulate the potential distribution difference between the training data and the final test data. Specifically, %we first conducted statistics on the recording time and location of each audio in the evaluation set, such as the city where the audio was recorded, the recording month, and the day-night distribution. Secondly, 
we first extract the Log-Mel filter bank (LMFB) features from audio files in both the evaluation set and the validation set, then calculate the statistical distribution of each audio feature, such as the mean and variance. 
%At the same time, we performed statistics on the validation set following the same steps.
%In the second step, we retained the samples in the validation set that belonged to the recording cities covered by the evaluation set and whose recording time periods highly overlapped with those of the evaluation set.In the second step, we further analyzed the distribution of their audio feature validation. We extracted the LMFB features of the validation set samples and 
In the second step, we use cosine similarity to quantify the differences between the feature distribution of the validation set and the evaluation set. A threshold of 0.9 is used: all samples with cosine similarity higher than 0.9 are considered to have a consistent distribution with the evaluation set and thus assigned to Valid-Hard, while the remainder are allocated to Valid-Easy. 
% Furthermore, since we can not augment the dataset to introduce diverse audio features, we adopt an alternative, simpler approach: randomly shuffling the recording time and location metadata of the audio to simulate varying data distributions, yielding Valid-Hard(x\%). We conduct experiments on 5\%, 10\%, 20\%, 50\%, 70\%, and 100\% of the Valid-Hard data, respectively. This division allows us to account for not only domain shifts between the training set and the final test set audio files but also the influence of spatiotemporal information on the model. 

Furthermore, considering that in real-world applications the spatiotemporal distribution of audio data may differ from the training domain, we randomly shuffle the recording time and location metadata of a portion of the Valid-Hard set. The proportion of shuffled data is denoted as x\%, referred to as Valid-Hard (x\%). Specifically, we set x to 5, 10, 20, 50, 70, and 100, in order to more comprehensively evaluate the impact of spatiotemporal distribution shifts on model performance.

%We regard the Valid-Hard as "hard samples" to test the generalization ability of our model on unseen data.
%To be specific, the distribution discrepancy mentioned here not only refers to differences in metadata such as the record time and location of audio recording, but more importantly, the distribution shift of audio features themselves caused by these metadata differences. In practical audio classification tasks, such distribution deviation of audio features can more directly affect the recognition effect of the model, thereby influencing the classification accuracy of the model.

%Specifically, Valid-easy consists of data that has a large distribution difference from the evaluation set and a small distribution difference from the training set, while Valid-hard has a small distribution difference from the evaluation set and thus is more similar to the evaluation set in distribution. We regard this part of data as "hard samples" to test the generalization ability of our model on unseen data.

% we split the validation data into two subsets: \textbf{Valid-Easy} and \textbf{Valid-Hard}. Specifically, the samples in Valid-Easy exhibit relatively small spatiotemporal distribution differences from the training data. In contrast, considering that in real-world applications the spatiotemporal distribution of data may differ substantially from that of the training set, we assign samples with larger distributional discrepancies to Valid-Hard.

\subsection{Evaluation Metric}
Following the challenge baseline, we evaluate the performance of the ASC system using accuracy (ACC) as the primary metric. Accuracy measures the proportion of correctly classified samples over the total number of samples, providing a straightforward and interpretable assessment of the model’s classification effectiveness across different scene categories.

\subsection{Training Details}
We first resample the audio recordings in the TAU UAS 2020 and CAS 2023 datasets to 44.1 kHz. All audio clips have a fixed length of 10 seconds. LMFB were extracted as audio features by using the Librosa~\cite{mcfee2015librosa} library with 2048 short-time Fourier transform (STFT) points, a 40ms Hann window, and a frame shift of 20ms. We apply 64 Mel-filter bands on the spectrograms and generate a feature tensor shape of 500 × 64 × 1. Dropout rate is set to 0.1. We train our model using the Adam~\cite{adam2014method} optimizer. The Batch size is set to 4 and the learning rate is set to 0.0001. All of our models are trained using the PyTorch toolkit~\cite{paszke2019pytorch} due to its ease of use and efficient support for rapid prototyping and debugging.
%because of its convenient capabilities.

We adopt a four-stage semi-supervised training framework that progressively improves model performance by effectively leveraging labeled and unlabeled data. Firstly, we train the ASCMamba model using our self-generated pretraining dataset to obtain a base pretrained model. In the second stage, we fine-tun the model using the labeled portion of the development dataset. During the third phase, we generate pseudo-labels for unlabeled development data based on the fine-tuned model from the previous step, identifying reliable data within this subset. 
Specifically, to identify reliable data for the second round of fine-tuning in the third stage, we implement a two-stage selection strategy as follows: first, guided by the confidence scores of pseudo-labels, the top 90\% data instances with the highest reliability are directly designated as reliable data. Second, for the remaining 10\% of data instances, ASCMamba and the improved SE-Trans are employed to generate pseudo-labels. Audio samples predicted to belong to the same scene category by both of 2 models are selected as reliable pseudo-labeled data. This strategy is designed to maximize the utilization of officially provided training data, thereby enhancing the model's generalization capability.

Finally, we perform another fine-tuning using both the real labeled and reliable pseudo-labeled data from the development set on the fine-tuned model, ultimately obtaining a final model that balances accuracy and generalization capability.

% Because of the size of the data from the challenge dataset is much smaller than that from the pre-training dataset. To ensure that the model sees the challenge data more frequently during fine-tuning, 
% To identify reliable data for the second round of fine-tuning in the third stage, we implement a two-stage selection strategy as follows: first, guided by the confidence scores of pseudo-labels, the top 90\% data instances with the highest reliability are directly designated as reliable data. Second, for the remaining 10\% of data instances, ASCMamba and the improved SE-Trans are employed to generate pseudo-labels. Audio samples predicted to belong to the same scene category by both of these two models are selected as reliable pseudo-labeled data. This strategy is designed to maximize the utilization of officially provided training data, thereby enhancing the model's generalization capability.
% the top 90\% of data instances is high-quality data, which are selected as reliable data.
\section{Results and Discussions}
\begin{figure}
\centering
\includegraphics[width=\linewidth]{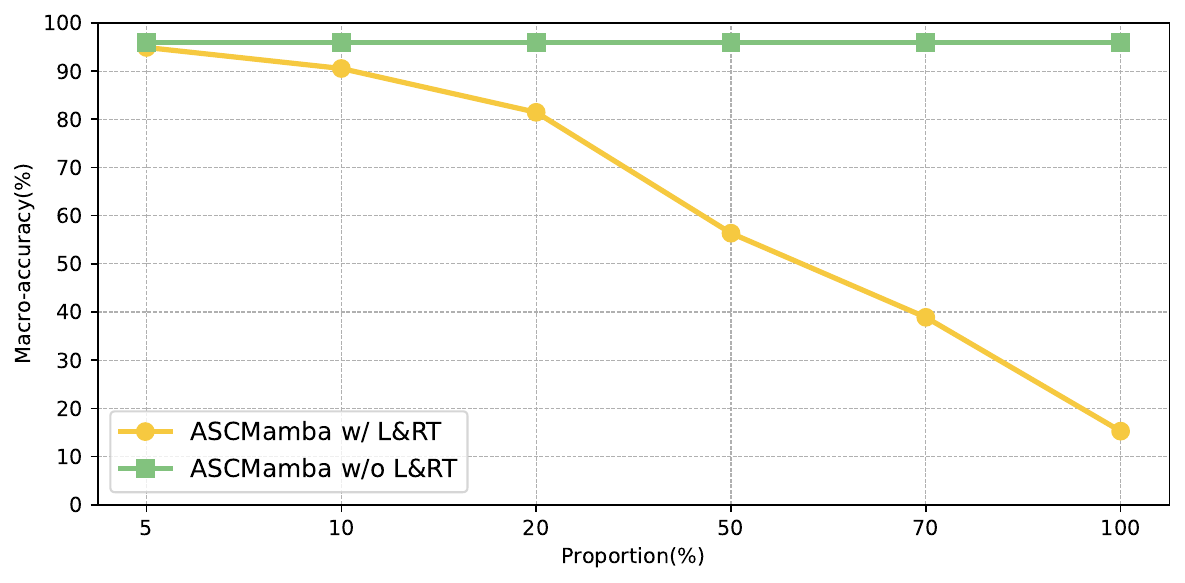}
\caption{The performance of the two systems after shuffling the Valid-Hard data at different proportions}
\label{fig:shuffle}
\end{figure}

To demonstrate the effectiveness of our model in audio classification tasks, we conduct the following ablation studies. 
%We evaluate two model variants alongside the baseline, assessing their performance on both Valid-Easy and Valid-Hard datasets. 
We evaluate the performance of the baseline and the 2 model variants separately on the Valid-Easy and Valid-Hard datasets.
The results are listed in Table~\ref{tab:result}. When evaluated on Valid-Easy, ASCMamba w/ L\&RT achieves the best performance with an average accuracy of 97\%. This strongly validates the model's ability to extract and learn audio signal features. In contrast, when tested on Valid-Hard, ASCMamba w/o L\&RT outperforms ASCMamba w/ L\&RT. Additionally, following the division method described in \ref{sec:validset creation}, we separately evaluate the performance of the 2 model variants on Valid-Hard when spatiotemporal information is randomly shuffled at different proportions, and the results can be found in Fig.~\ref{fig:shuffle}. 

\begin{figure}
\centering
\includegraphics[width=\linewidth]{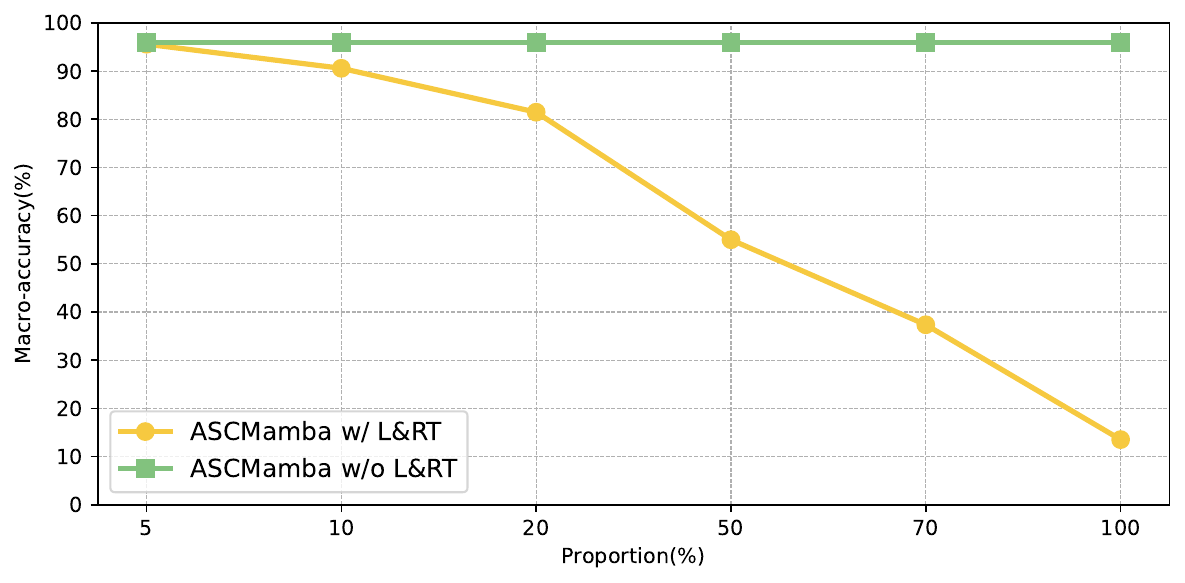}
\caption{The performance of the two systems with the gradual increase in the proportion of unseen locations within the validation set
}
\label{fig:unseen city}
\end{figure}

In addition to randomly shuffling spatiotemporal information, we also notice that the evaluation set contains locations unseen in the training set, and thus we conduct further experiments: we select different proportions of data from the validation set. The proportions of these selected data are the same as those of Valid-Hard, and randomly change the location to those seen in the evaluation set but not in the validation set, such as \textit{Nanchang}, \textit{Shenyang}, \textit{Guangzhou}, \textit{Changchun}, \textit{Tianjin}, and \textit{Taiyuan}. The experimental results are visualized in Fig.~\ref{fig:unseen city}.

Results on Fig.~\ref{fig:shuffle} and Fig.~\ref{fig:unseen city} show that, when the spatiotemporal distribution of the test data differs from the training domain, or when unseen locations appear in the test set, ASCMamba w/ L\&RT consistently underperforms ASCMamba w/o L\&RT. 
Furthermore, greater distributional discrepancies lead to a more substantial degradation in the performance of ASCMamba w/ L\&RT relative to its counterpart.

We attribute this to the reliance of ASCMamba w/ L\&RT on temporal and positional information, which arises from its extensive utilization of multi-modal information. 
This also implies that when external multi-modal information is ambiguous, it is preferable to solely utilize audio-specific features for analysis; whereas when multi-modal information is clear, employing multi-modal fusion can significantly enhance the model's discriminative performance.

%To further validate our hypothesis, we designed two additional experiment: Firstly, we conducted experiments following the division method of Validation Sets Creation. 

%Furthermore, in order to study the influence of seen/unseen locations on the experimental results, we also selected different proportions of data from the original validation set. The proportions of these selected data was the same as those of Valid\_Hard\_1, and randomly changed the location to those seen in the evaluation set but not in the validation set, such as Nanchang, Shenyang, Guangzhou, Changchun, Tianjin and Taiyuan, so as to generate Valid\_Hard\_2. The experimental results, visualized in Figure~\ref{fig:shuffle} and Figure~\ref{fig:unseen city}, confirm our hypothesis.

% \begin{table}
%     \centering
%     \caption{Shuffling the temporal and positional information of Valid-Hard data at different proportions}
%     \begin{tabular}{c|cccc}
%     \midrule
% System & 10\% & 20\% & 30\% & 50\% \\ \hline
% ASCMamba w/o L\&RT & 90.55 &81.45 &73.45 &56.36\\
% ASCMamba w/ L\&RT & 96.00 &96.00 &96.00 & 96.00\\ \hline
%     \end{tabular}
%     \label{tab:valid-hard}
% \end{table}

The evaluation results of each team on the official blind test set in Table~\ref{tab:evaluation results}. It is clearly evident that our system achieves a 6.2\% improvement over the official baseline system and outperforms all systems submitted by other participating teams, securing first place in this competition. Rank-2 System~\cite{Rank-2} proposes a city-separated cross-validation scheme to evaluate the model's generalization capability on unseen locations. It employs low-frequency feature constraints to mitigate overfitting risks and replaces max-pooling with average pooling for aggregating temporal frame information. However, it has limitations such as restricted applicability due to the low-frequency assumption, under-utilization of multi-modal information, and unknown robustness to extreme data cases. Rank-3 System~\cite{Rank-3} employs an attention-based ResNet as the audio backbone network, which effectively captures key features in the time-frequency domain. However, this system fails to propose an intrinsic mechanism to address the model's heavy reliance on positional information, which may hinder its generalization to unseen locations.

%It is clearly observed that our model exhibits a substantial improvement over the models proposed by other teams. Furthermore, it achieves an approximate 6.2\% improvement compared to the baseline, which strongly validates the superiority of our proposed model.

%The experimental results on validation sets are presented in Table~\ref{tab:result}. On Valid-Easy, the accuracy rate of model classification has reached 97\%.

%For Valid-Hard, ASCMamba without city and time performs better, which can prove its sensitivity to spatio-temporal information. If the distribution of spatio-temporal information between the training data and the final test data is inconsistent, it will lead to a deterioration of the model's performance. Therefore, we use ASCMamba w/o L\&RT as the final submitted system.

% It can be seen that the proposed system achieves superior classification performance across all scene classes, with an average precision 7.34\% higher than that of the baseline system, demonstrating the effectiveness of our approach. Notably, the classes \textit{Airport}, \textit{Bar}, \textit{Bus}, \textit{Construction Site}, \textit{Republic Square}, \textit{Restaurant}, and \textit{Urban Park} all achieved 100\% accuracy, further demonstrating the efficacy of the proposed ASCMamba model.

\begin{table}
    \centering
\caption{The evaluation results of each team on the official blind test set}
    \begin{tabular}{cc}
        \midrule
        System& Score(Macro-accuracy) \\ \hline
         Rank-1 (ASCMamba w/o L\&RT) & \textbf{64.4\%} \\ 
         Rank-2 System & 62.8\% \\ 
         Rank-3 System & 61.3\% \\ 
         Rank-4 System & 58.6\% \\ 
         SE-Trans (Baseline) & 58.2\% \\ \hline
    \end{tabular}
    \label{tab:evaluation results}
\end{table}

\section{CONCLUSION}

In this paper, we present our approach to tackle the ASC task of the APSIPA ASC 2025 Grand Challenge. In detail, we propose a novel architecture named ASCMamba, which uses a DenseEncoder to extract local and hierarchical features, and applies a Dual-path Mamba block for sequence modeling. In addition, we employ a two-step mechanism to generate reliable pseudo-labels for unlabeled data with low confidence. Experimental results in demonstrate that ASCMamba exhibits distinct advantages in fusing multi-modal information.
On the official blind test dataset, the proposed ASCMamba outperforms the existing baseline system by 6.2\% in macro ACC, and achieves the first rank among all participating teams, which sufficiently demonstrate the effectiveness of the proposed approach in acoustic scene classification tasks. 

\footnotesize
\printbibliography

\end{document}